# DATA PREPROCESSING FOR PARAMETER ESTIMATION.
## AN APPLICATION TO A REACTIVE BIMOLECULAR TRANSPORT MODEL

Cuch, Daniel A. ‡;  Rubio, Diana†;  El Hasi, Claudio D. ‡

*‡Instituto de Ciencias - Universidad Nacional de Gral Sarmiento, J.M. Gutierrez 1150, Los Polvorines, Buenos Aires, Argentina*
*†Centro de Matemática Aplicada, ECyT - Universidad Nacional de Gral San Martín. M. De Irigoyen 3100, 1650 San Martín, Buenos Aires, Argentina, drubio@unsam.edu.ar*

**Abstract**

In this work we are concerned with the inverse problem of the estimation of modeling parameters for a reactive bimolecular transport based on experimental data that is non-uniformly distributed along the interval where the process takes place.

We proposed a methodology that can help to determine the intervals where most of the data should be taken in order to obtain a good estimation of the parameters. For the purpose of reducing the cost of laboratory experiments, we propose to simulate data where is needed and it is not available, a PreProcesing Data Fitting (PPDF).We applied this strategy on the estimation of parameters for an advection-diffusion-reaction problem in a porous media. Each step is explained in detail and simulation results are shown and compared with previous ones.

1. INTRODUCTION

The number of studies on environmental issues, related to reactive transport has increased in the past few years, showing the relevance of this subject (see, for example: Kourakos and Harter, 2014; Edery et al., 2013 and references therein). Air and water masses behave as mobile reservoirs, and are responsible of bulk advective pollutant transport. Analysis of solute flow dynamics takes into account that, in actual situations, there is also a spreading or mixing phenomenon associated to the advective movement: this dispersion spread out sharp fronts, resulting in the dilution of the solute.Interaction of the pollutants with the solid phase of soil or particulate matter through sorption processes results in retarded fronts and changes in concentration (Logan, 1999). Also, the transport of reacting species is affected by the changes produced in the chemical composition of the environment. Hence, a modeling tool is needed in order to achieve a deeper understanding of the phenomena that can be applied to a large number of applications, such as waste disposal management, drinking water supply protection and environmental remediation.

Conceptual models of soil processes are very useful to predict and/or understand the movement of different species in the environment (Huyakorn and Pinder, 1983). An important number of the models are deterministic, based on conservation laws for mass, energy and momentum. Subsurface transport processes are ruled by Darcy's law and conservation of mass. The main objective is to calculate concentration of chemicals dissolved in water as a function of space and time. In a large number of cases, advection-dispersion-reaction models (ADRE) are based on considering the porous media as continuous phase, averaging the concentration values over many pore spaces (Bear, 1988). The models are designed to reproduce observations at macroscopic level as results of processes that take place at a microscopic scale.

When flow processes with multiple reactive species are studied, it is important to consider the nature of reactions and the role of fluctuations at small scale (Porta et al., 2012; Edery et al., 2013; Chiogna and Bellin, 2013). In the smaller ranges the fluid velocity is never homogeneous in space, and the continuum hypothesis loses validity. While the equations at Darcy´s scale are based on a continuum hypothesis, averaging over an elevated number of pores, the reaction dynamics are governed by poral scale processes (Kapoor et al., 1997). In many cases, a suitable approach is to analyze simplified schemes, such as considering one-dimensional flow linked to bimolecular reactions (Kapoor et al., 1998). The differences between empirical and numerical data depend on how these processes are modeled (Raje and Kapoor, 2000; Gramling et al., 2002; Sanchez Vila et al., 2010).This approximation leads to big estimation errors. Usually the parameters of the simulations are determined in batch reactors and supposed to be valid in transport processes. Mathematical models that describe diffusive-reactive processes more accurately, incorporating poral scale effects, the so called *segregation,* have been developed (Meile and Tuncay, 2006; Rubio et al., 2008). New transport models have been studied considering how the features occurring at poral scale can be reflected at mesoscale (Cuch et al., 2009, Porta et al., 2012).

There exists some discrepancy between the rates of reaction at the field and laboratory. Moreover, many difficulties arise in adjusting the results among different scales when averaging over many pores to obtain the continuum equation. Assumptions made are to assume that the poral scale is much smaller than the average volume, where the mixture takes place instantaneously and completely. These assumptions are not reasonable if the fluctuations at poral scale are large (Cirpka, 2004);in this case the concentration gradients play an important role. Furthermore, an over-estimation of the reaction products is usually obtained. Although the formulation of the continuum based on ADRE has restrictions (Edery et al., 2013), and other methods such as Particle Tracking are used to try to link the effects between the different scales (Chiogna and Bellin, 2013), the ADRE model is useful for comparison as modeling tool in reactive transport processes.

Here we focus on studying transport in porous media considering a macroscale (continuum) approximation modeled by a system of partial differential equations that link different effects: advection, dispersion and reaction. Fluctuations that occur at microscale (poral level) will be taken into account considering upscaled/effective parameters.

Usually mathematical modeling of a physical process involves an inverse problem that consists in estimating one or more parameters of the model based on experimental data (observations) (Tarantola, 2005). In recent years the study of the inverse problem took much interest in applications arising from different disciplines: engineering, biology, economics and even medicine (see for example Brown and Jais, 2011, Andrle et al 2011). Essentially, it is an optimization problem that consists in finding estimated values of modeling parameters for which the simulated solution accurately fits the available experimental data. The proposed mathematical model is critical and the optimization method chosen is also important for this purpose. (Blocken and Gualtieri, 2012). Once the model is established, the parameters are usually estimated by minimizing the square errors between the simulated values and the experimental data. Generally, experimental data are dispersed and are likely to provide information that is not enough for the correct modeling, whereby a skilled data preprocessing may be beneficial for the optimization process.

In this work we considera one-dimensional ADRE model that adequate the reaction rate in the transport equation incorporating the segregation term as an effective rate that depends linearly on a free parameter and consider the diffusion by another independent parameter. Furthermore, we propose a simple way to reduce the computing time by fitting the experimental data by a smooth function. Also, we introduce a step variation scheme that takes into account the regions where the shape of the data curve vary more rapidly according to the fitting function. Numerical simulations were made and the segregation and dispersion parameters were estimated by analyzing the production profile and the mass of product.

2. MATHEMATICAL MODEL FOR THE TRANSPORT PROCESS

It is necessary to understand the dynamics of the process being studied in order to obtain a mathematical model that properly describes the problem. In this case the elements to be considered are the porosity, the flow rate, the tensor of dispersion-diffusion, the processes at the interface and the reaction rate of interacting species.

The equation used to model solute transport in porous media is generally a non linear differential equation in partial derivatives of second order of parabolic type,

$$\frac{\partial(\phi c(t,x))}{\partial t} + \nabla[(\mathbf{V}\phi c(t,x) - \mathbf{D}\cdot\nabla c(t,x))] = S(t,x), \qquad (t,x) \in \Omega \tag{1}$$

being $\phi$ the medium porosity, c the solute concentration, $\mathbf{V}$ the flow rate, $\mathbf{D}$ the dispersion-diffusion tensor of solute and $S$ a source term whose shape depends on the problem under study (adsorption, degradation, reaction, etc.). Finally $\Omega$ is the region where the process takes place.

For simplicity, we assume that the reactives move along an enclosure of uniform section in which the dynamics along the direction of displacement is the only relevant one, so that it can be assumed that the transport process is one-dimensional. Although the velocity $\mathbf{V}$, dispersion $\mathbf{D}$, and the porosity $\phi$ can be space dependent, given the assumption of homogeneity they may be considered constant throughout $\Omega$. Therefore, the reaction process can be described by a set of equations of the type:

$$\frac{\partial c_i(t,x)}{\partial t} + V\frac{\partial c_i(t,x)}{\partial x} - D^*\frac{\partial^2 c_i(t,x)}{\partial x^2} = S^*(t,x), (t,x) \in \Omega, \; i = 1,2,3 \tag{2}$$

where $c_i$ are the solutes concentration, $D^* = D/\phi$, $S^* = S/\phi$ and $\Omega = [0,L]x[0,T]$.

This equation itself is quite complex to solve numerically and the method used for it depends on the relationship between the characteristic times of advection, diffusion and reaction ($t_A, t_D$ and $t_R$, respectively). As it is known, the advective *Damköhler number* ( $Da_A \equiv {t_A}/{t_R}$ ) indicates whether the advection is the dominant process compared with reaction. When $Da_A << 1$ experiments can be simulated with standard numerical techniques (Press et al., 1992). Meanwhile, if $Da_A >> 1$ the reaction is faster than the advection, and it is difficult to numerically simulate because the integration step must be smaller than $t_R$. In this case the reaction time step is several orders of magnitude smaller than the time step needed to integrate the advective process, requiring too much computing time. Wheeler and Dawson (Wheeler and Dawson, 1987) proposed an *Operator Splitting* method, were each integration stage is done in two steps: the first solves the advection-diffusion equation without the reactive term, dealt with in the next step.

There are several integration methods, we choose an integration step $\Delta t$ splitted in two successive steps. In the first step we consider only the advection process, and an integration step $\Delta t_1$. Afterwards, the numerical result is used to integrate the process considering only reaction, with a time step $\Delta t_2$ where $\Delta t_1 >> \Delta t_2$ and $\Delta t = \Delta t_1 + \Delta t_2$. These steps we may described by the following equations:

$$\frac{\partial c_i}{\partial t} + V\frac{\partial c_i}{\partial x} - D^*\frac{\partial^2 c_i}{\partial x^2} = 0 \tag{3}$$

$$\frac{dc_i}{dt} = -S^* \tag{4}$$

Eventhough the physical process is unique, one may think that it is necessary a $\Delta t_1$ time to mix solutes, then react during a shorter time $\Delta t_2$ giving products; after that, a new period $\Delta t_1$ is needed to mix and homogenize the reactants, and so on.

Because of the characteristics of the problem, appropriate time steps must be chosen to solve iteratively equations (3) and (4). An arbitrary integration time step $\Delta t_2$ may yield wrong numerical results that would not accurately fit the experimental data. A relationship between time steps that depends in a simple way on the characteristic times (equation 5) has been proposed by Rubio et al. (Rubio et al., 2008):

$$\Delta t_2 = \Delta t_1 / Da_A \tag{5}$$

Note that the ratio between the two intermediate time steps is given by the Damköhler number, indicating the relative weight of each term in the whole process. The equations (3) and (4) are solved by turns and the numerical integration must be repeated until the final time is achieved.

For the numerical solution of the equations we consider a finite difference scheme centered in space and forward in time, which guarantees a first order accuracy in time and second order in space (Rubio et al., 2008). Appropriated initial and boundary conditions are set for the process under consideration.

3. NUMERICAL MODELING

In the literature one can find an important number of publications considering this type of problems. Some of them are focus on some theoretical aspects (Kapoor et al., 1997) and others are concerned on the experimental issues (Kapoor et al., 1998, Gramling et al., 2002, Raje and Kapoor, 2000). In this paper we focus on numerical experiments that could provide on one hand a simple and effective manner to manage the data, and by the other an effective way of modeling the problem of segregation.

We consider two reactive solutes A and B in a transport flux that produces C. Assuming a stationary absorption process between the solid and liquid phases, and homogeneity of the reactive substances, the process may be considered as a bimolecular transport.

For a bimolecular reaction process we assume that

$$S = \Gamma c_1 c_2 \tag{6}$$

where $\Gamma$ is the reaction rate and $c_1, c_2$ are the concentration of the reactants A and B respectively. Usually mean concentrations $\bar{c}_i$ are considered instead of point values $c_i$, being the fluctuation $c_i'$ the difference between them, that is, $c_i' = c_i - \bar{c}_i$.

In the flow of reactive solutes the continuum approach can produce erroneous results since reactants are considered homogenized in Darcy scale but they are no perfectly mixed in poral scale (Gramling et al., 2002 and Kapoor and Raje, 2000). In this paper we use a one-dimensional model that changes the rate of reaction in the transport equation incorporating an effective reaction rate (Meile and Tuncay, 2006, Rubio et al., 2008), which linearly depends on a free parameter. This parameter is determined from experimental data by minimizing the sum of squared residuals. Equation (6) must be replaced by (Kapoor et al. 1997)

$$S = \Gamma \overline{c_1 c_2} = \Gamma(1+s)\overline{c_1}\overline{c_2} \tag{7}$$

being $s$ the segregation factor. Changes in concentration at poral scale can be modeled with this correction factor which takes into account the macroscopic gradients of concentrations and a parameter to determine (Meile and Tuncay, 2006):

$$s = \frac{\overline{c_1' \cdot c_2'}}{\overline{c_1} \cdot \overline{c_2}} \approx \frac{\alpha}{\phi} \frac{\overline{\nabla c_1} \cdot \overline{\nabla c_2}}{\overline{c_1} \cdot \overline{c_2}} \tag{8}$$

Finally, the equation (7) takes the discretized form

$$S_{i,j} = \Gamma \left[ 1 + \frac{\alpha}{\phi} \frac{(c_{1i,j+1} - c_{1i,j-1})(c_{2i,j+1} - c_{2i,j-1})}{(2\Delta x)^2 \, c_{1i,j} c_{2i,j}} \right] c_{1i,j} c_{2i,j} \tag{9}$$

In practice, a certain number $M$ of experimental data are available and they might not be enough or they might not be taken at the region (or time instants) of interest. Notice that data contain measurement errors and an appropriated (no interpolant) fitting function that best approximate observed data, may be considered. Afterwards, the resulting fitting curve could be used to obtain approximated data when it is needed and the experiment cannot be repeated.

The idea of this work is to use values of the fitted function at points of interest as if they were experimental data. A study or analysis of the behavior of the particular process of study can help to choose the fitting curve to be used and the number and location (in time or space) of the simulated data.

Observation data at the intervals where the second derivative achieves its highest (absolute) values would provide valuable information for the parameter estimation, since that intervals characterize the distribution shape.

3.1 The PPDF strategy

The procedure we propose here (PPDF) can be described by the following set of steps:

1) Fit the available data by a function *g(x)*. This can be done with the MATLAB function *fit* using one of the library models present. As we mentioned above, a Gaussian function is considered in this work.

2) Define a uniform grid G in [0,L] and find the points $x_{mG} \in G$ where the absolute values of the second derivative of the fitting function are highest,

$$x_{mG} = argmax\{|g_{xx}(x)|, x \text{ in } G\}$$

3) Define a new grid $G^*$ by refining G around the points $x_{mG}$.

4) Numerically calculate simulated data $c_j = c_j(x_j)$ at the new grid points $x_j \in G^*$.

5) Find the estimated value $\hat{\alpha}$ of $\Omega$ that minimizes the square error

$$\hat{\alpha} = \arg\min \sum_{j=0}^{N} \frac{\left(d_j^{PPDF} - c_j\right)^2}{N} \qquad (10)$$

where $d_j^{PPDF}$ are the values of the fitting function at $G^*$ and $c_j = c_j(x_j)$ are the simulated values calculated in the previous step.

The procedure described above can be applied to a wide range of situations, even in cases with a large variability in the distribution of experimental data. In particular, in the cases that we discuss later, the profile data has areas of very rapid change. In situations like this one, some details of the profile can be lost when using a large integration step. On the other hand, the computation becomes very slow if the step is too small. Instead of using adapting steps procedures, we develop a simpler strategy that considers a shorter integration step in areas where the rate of change of the concentration of product is most significant.

4. APPLICATIONS AND DISCUSSION

In the process that we are considering, segregation phenomenon occurs, whereby the prediction of continuous models generally do not match the experimental data (Ederly/Chiognia and references therein). In this section, we apply the proposed strategy to Gramling *et al* and Raje and Kapoor works.

Because of the shape of the concentration data profile that we are analyzing, a Gaussian function seems to be the most appropriated for fitting the experimental data. We have also estimated the diffusion coefficient (*D*), since it is not clear that its value is the one reported for nonreactive experiments (Edery et al, 2013 and references therein). Hence, the algorithm described in the previous section was implemented for the problem of the estimation of both, the segregation and the diffusion coefficient of the transport process.

As a first example we consider the process described in Gramling *et al.*, 2002 and their experimental results. The concentration of product ($c_3$) at each point *x* of the column is reported in four fixed instants of time, and the values of the characteristic constants of the experiment: reaction constant ($\Gamma$), tube length (L) and porosity ($\phi$), are given. The results of the simulations are shown in Figure 1 wherein the concentration of product is observed for the incoming solution for a particular experiment conducted by Gramling *et al.* corresponding to a flow rate of V = 0.0121 cm/s. The red dotted lines correspond to the experimental data, the dashed lines corresponds to an analytical solution (Gramling *et al.*) without considering the segregation while the blue solid line is the result of our simulations obtained after performing a previous Gaussian fit and locally refine the grid, as it was explained in the previous section.

It can be seen that the simulated curve fit well the experimental results and provide a good estimation of the concentration profile for $c_3$. Similar results are obtained in all cases reported in Gramling et al.

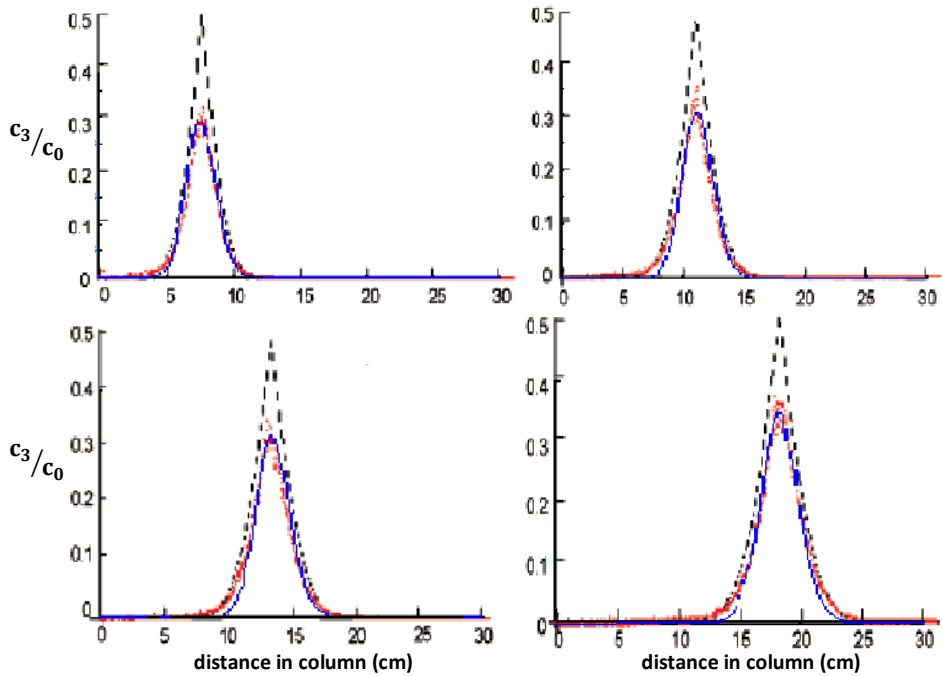

**Figure N°1**. Product concentration with respect to the initial value along the column for time instants T = 619 s, 916 s, 1114 s, 1510 s. V = 0.0121 cm/s.

---- experimental data (Gramling et al., 2002)   ---- **analitical solution** (Gramling et al., 2002)   ----- our numerical solution

Table 1 shows the estimated values $\hat{\alpha}$ and $\widehat{D}$ for the parameter $\alpha$, related to the segregation factor, and the dispersion coefficient D, respectively.

We estimate independently both parameters for each one of the four times reported in Gramling's work, in order to analyze the consistency of our results. Also we have observed good results when we use one specific time to estimates the values of the parameters and then simulate the process for the other three times. This second strategy was performed by Chiogna and Bellin.

|                   | 619 s  | 916 s  | 1114 s | 1510 s |
|-------------------|--------|--------|--------|--------|
| $\hat{\alpha}(cm/s)$ | 0.131  | 0.166  | 0.158  | 0.164  |
| $\hat{D}(cm^2/s)$ | 0.0013 | 0.0012 | 0.0011 | 0.0012 |

**Table N°1**. $\hat{\alpha}$ and $\hat{D}$ for T = 619 s, 916 s, 1114 s, 1510 s. V = 0.0121 cm/s.

As before, we assume that both, the parameter $\alpha$ of the segregation factor and the dispersion coefficient $D$, are constant over time. The estimated values shown in Table 1, for different sets of data (taken at different times) might indicate a dependency on time for both parameters. Nonetheless since the dispersion is small, we might think they are constant in time for this problem (with the limitations imposed by the small number of cases) and equal to $\bar{\alpha} = 0{,}155$ and $\bar{D} = 0{,}0012$.

The estimated value $\hat{D}$ for the coefficient of dispersion is lower than that determined in nonreactive experiment ($D = 0.00175 \frac{cm^2}{s}$) in Gramling *et al.* which would give a narrower concentration profiles (see Rubio et al., 2008).

Another indicator of the reliability of the model is their ability to reproduce the total amount of mass produced. We calculate the mass production considering the molecular weight of the product, and the concentrations for the experimental case under different approximation. The results are shown in Figure 2, .It can be observed that, although the simulation using the PPDF methodology introduced in this work (using Gaussian fit) give values a little lower than the experimental ones, they provide a good approximation.

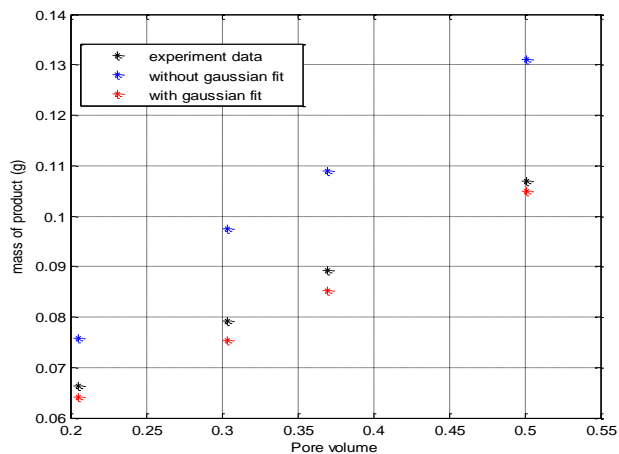

**Figure N°2**. Total mass of the product for T = 619 s, 916 s, 1114 s, 1510 s., for the experimental data (Raje and Kapoor, 2000) and different numerical simulations. V = 0.0121 cm/s.

Now, we consider other experimental settings reported in Gramling et al., and the results are shown in Table 2. It can be observed an increased value for the velocity dispersion, which is an expected result, since at low speeds the dispersion is proportional to the velocity ($D = \lambda V$), being dispersivity $\lambda$.

| Q | 0.0121 cm/s | 0.0832 cm/s | 0.67 cm/s |
|---|---|---|---|
| $\hat{\alpha}(cm/s)$ | 0.155 | 0.196 | 0.23 |
| $\hat{D}(cm^2/s)$ | 0.0012 | 0.0116 | 0.13 |

**Table N°2**. $\hat{\alpha}$ and $\hat{D}$ for different velocities: 0.0121 cm/s, 0.0832 cm/s and 0.67 cm/s.

The estimated value $\hat{D}$ for the coefficient of dispersion for V= 0,0832 cm/s and V = 0,67 cm/s are also lower than that determined in the nonreactive experiment ($D = 0.0145 \frac{cm^2}{s}$ and $D = 0.175 \frac{cm^2}{s}$, respectively) in Gramling *et al*. being the differences about 25%.

Another example is based on results published by Raje and Kapoor, 2000. In this case the concentration of the product is measured at the end of the column (*x* fixed), as a function of time. The experiment is performed at two different flow rates. The setup and the values of the parameters of interest are given in (Raje and Kapoor, 2000)

Figure 3 shows the results for the case of an initial concentration of 0.5 mM for both reactants A and B, and a flow rate of 0.096 cm/s. The squares correspond to the experimental data, blackline show the results without considering the segregation effects and blue line correspond to the simulated result obtained using the PPDF strategy..

Figure 4 shows the analogous result for an initial concentration of 0.25 mM for reactants A and B and a flow rate of 0.07 cm/s.

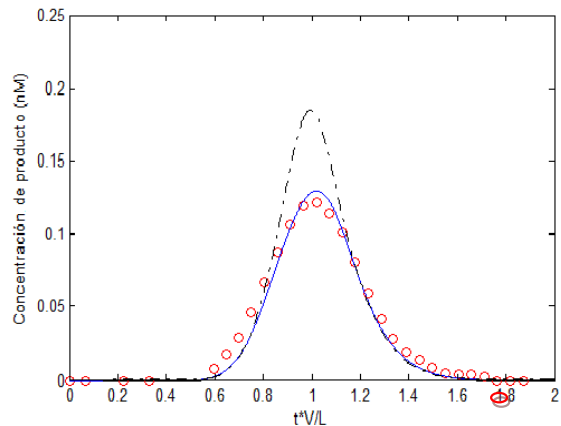

**Figure N°3**. Product concentration at the end of the column as a function of time. Initial concentration for reactants: 0.5 mM, flow velocity: 0,096 cm/s.
-.-.-.-numerical solution without gaussian fit and parameters (Raje and Kapoor, 2000)  ⊖⊖ experimental data (Raje and Kapoor, 2000)   ------.our.numerical solution

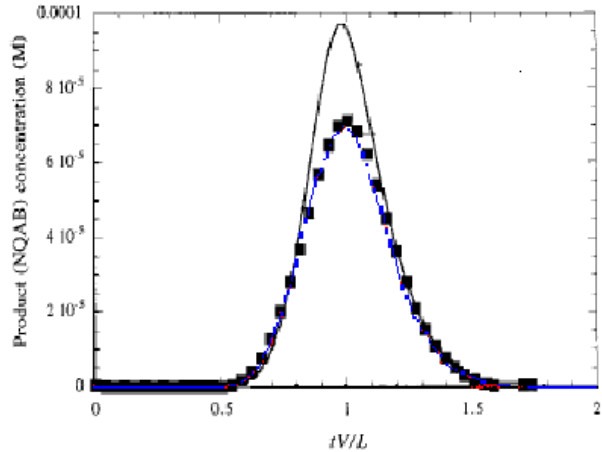

**Figure N°4**. Product concentration at the end of the column as a function of time. Initial concentration for reactants: 0.25 mM, flow velocity: 0,07 cm/s.
■■■■■ experimental data      ---------- analitical solution      ---------- ..numerical solution

In both cases our numerical simulations fit very well the experimental data. The estimated values $\hat{\alpha}$ and $\hat{D}$ obtained for the segregation coefficient $\alpha$ and for the dispersion coefficient D are shown in Table 3.

| Q | 0.07 cm/s | 0.096 cm/s |
|---|---|---|
| $\hat{\alpha}(cm/s)$ | 2 | 1.81 |
| $\hat{D}(cm^2/s)$ | 0.025 | 0.034 |

**Tabla N°3**. $\hat{\alpha}$ and $\hat{D}$ for an initial concentrations of 0.25 nM and 0.5 mM for the reactants A and B.

The estimated value $\hat{D}$ for the coefficient of dispersion is a little higher than that determined in the nonreactive experiment ( $D = 0.023 \frac{cm^2}{s}$ and $D = 0.032 \frac{cm^2}{s}$, respectively) in Raje and Kapoor, the differences are about 7 %. Once more we note that the dispersion coefficient increases with speed, as it was expected.

In both experiments we can observe the dependency of the dispersion coefficient on speed. The values for $D$ obtained for the Gramling case are smaller than the ones of the experiments with nonreactive flow, reported in the original paper. For the ADRE model it is assumed that the transport process obeys the Fick's Law ($j = -D\,\nabla c$). As the reaction time is several orders of magnitude shorter than that of the advection and dispersion in the region of the reaction front, reactives are quickly consumed resulting in very large concentration gradients, this could explain why lower values of $D$ are obtained. In the other hand, in the experience of Raje and Kapoor, all characteristic times are of the same order of magnitude while the gradients are lower, which could explain why the value for D is similar to that determined in the non-reactive experiment.

These observations regarding the gradients of the reactants are in accordance with equation (8). The gradients, and its product, are greater in Gramlings than in Kapoor, which might explain why the parameter α is smaller in the first case

5. CONCLUSIONS

It is often difficult to have experimental data in the region of interest. They might be either scarce or not enough in the areas where the phenomenon to study present great variations or contain more information about the phenomenon. By means of a simple method that generates a smooth curve that approximates the experimental data, we can get "simulated data" that allow us to reproduce the experiments and, for instance, build a appropriated space discretization reducing integration step only in areas of interest. This methodology, that we have called PPDF, is simple, fast and the results presented here indicates its efficiency.

Furthermore, the use of macroscopic models (continuous), although it has its limitations, it may be useful to analyze, quickly and easily, a number of phenomena. Here we use the ADRE approach to analyze experimental results, showing discrepancies between the model and the experiments. However the inclusion of effective parameters combined with the methodology presented here for pre-processing the experimental data, appears as a useful tool to study the problems, providing information that is extremely useful and is consistent with what is expected.